\documentclass[reprint,aps,showkeys,superscriptaddress,
 prx,amsmath,amssymb,bm
]{revtex4-2}
\usepackage{graphicx}
\usepackage{subfiles}
\usepackage{algorithm,algpseudocode}
\usepackage{float}
\usepackage{multirow}

\begin{document}

\preprint{ASF and filtering}

%– Confidential Draft – Do NOT Distribute!!

\title{Adaptive folding and noise filtering for robust quantum error mitigation}

\author{Kathrin F. Koenig}
\email{kathrin.koenig(at)iaf.fraunhofer.de}
\affiliation{
 Fraunhofer Institute for Applied Solid State Physics, Tullastr. 72, 79108 Freiburg, Germany
}
\affiliation{Department of Sustainable Systems Engineering, University of Freiburg, Emmy-Noether-Str. 2, 79110 Freiburg, Germany}

\author{Finn Reinecke}
\affiliation{Department of Computer Science, University of Freiburg, Georges-K\"ohler-Allee 51
79110 Freiburg, Germany}

\author{Thomas Wellens}
\affiliation{
 Fraunhofer Institute for Applied Solid State Physics, Tullastr. 72, 79108 Freiburg, Germany
}

\date{\today}

\begin{abstract}

Coping with noise in quantum computation poses significant challenges due to its unpredictable nature and the complexities of accurate modeling. This paper presents noise-adaptive folding, a technique that enhances zero-noise extrapolation (ZNE) through the use of adaptive scaling factors based on circuit error measurements. Furthermore, we introduce two filtering methods: one relies on measuring error strength, while the other utilizes statistical filtering to improve the extrapolation process.
Comparing our approach with standard ZNE reveals that adaptive scaling factors can be optimized using either a noise model or direct error strength measurements from inverted circuits. The integration of adaptive scaling with filtering techniques leads to notable improvements in expectation-value extrapolation over standard ZNE. Our findings demonstrate that these adaptive methods effectively strengthen error mitigation against noise fluctuations, thereby enhancing the precision and reliability of quantum computations.
\end{abstract}


\maketitle

\section{\label{sec:introduction}Introduction}

Error mitigation in quantum computing aims to reduce computational errors without requiring full quantum error correction, which is resource-intensive and not yet feasible due to high error rates and limited number of physical qubits.

Zero-noise extrapolation \cite{Temme2017, Li2017, Endo2018} (ZNE) is a foundational technique for mitigating errors in noisy intermediate-scale quantum (NISQ) devices, enabling the estimation of noiseless quantum observables by analyzing results from circuits with amplified noise. Traditional approaches, such as digital ZNE \cite{He2020, Giurgica2020}, rely on the assumption that noise strength scales proportionally with an applied noise-scaling factor \cite{majumdar2023, Schultz2022}, while inverted-circuit ZNE (IC-ZNE) \cite{Koenig_2024} measures directly the error strength via an inverted circuit. Due to the possibility of erroneous results, it is not possible to ascertain their quality. In this regard, IC-ZNE can be instrumental, as it quantifies the error strength of a circuit. This information can be used to eliminate computations with high error strengths compared to other runs. To date, there is 
%tw an absence 
a lack
of documented filtering processes that are autonomous of the algorithm and that consider solely the outcomes of multiple executions.

The efficacy of Zero Noise Extrapolation (ZNE) can be improved using noise-aware folding, as proposed in \cite{hour2024}. Standard ZNE assumes uniform error across gates, which is inaccurate on hardware with non-uniform noise. Noise-aware folding uses calibration data to redistribute noise more effectively, adjusting circuits to balance error rates across logical qubits. This involves analyzing calibration data to tailor folding and align noise scaling with hardware-reported distributions. However, calibration data may be outdated due to temporal fluctuations, potentially leading to incorrect scaling and reduced mitigation effectiveness.

Despite the widespread application and development of ZNE as a quantum error mitigation technique, there has not been a lot of effort of research exploring the integration of filtering methods or additional post-processing steps to enhance ZNE's performance. While ZNE has been extensively studied and improved upon, including advancements in noise amplification methods \cite{pelofske2025, Schultz2022} and extrapolation techniques \cite{Halder2023}, the potential benefits of applying filtering or sophisticated post-processing to ZNE results remain largely unexplored. While Ref.~\cite{Kim2023_util}(supplementary material) employs a statistical filtering approach to improve error mitigation, a more systematic evaluation of its assumptions and applicability may provide further insights.

In order to overcome limitations of existing zero-noise extrapolation (ZNE) methods, this work introduces several techniques. These include: adaptive error scaling, which dynamically adjusts extrapolation based on directly measured real-time noise, and selective filtering, which removes high-error results from measurement data, thereby preventing faulty outcomes from skewing extrapolation. Notably, we also take into account the possible existence of a secondary Gaussian distribution to account for rare noise events and transient hardware instabilities, as well as the systematic rejection of runs associated with this secondary distribution. These techniques significantly advance ZNE's reliability and robustness.

We validate this adaptive framework using three benchmark algorithms, the Hadamard-Ladder circuit, Grover’s search \cite{grover1996} and the Harrow-Hassidim-Lloyd (HHL) \cite{Harr2009} algorithm, on superconducting quantum hardware. Our results demonstrate improvements of 23.65\% for adaptive scaling factors (ASF) and up to 29.81\% by separating faulty data and ASF in mitigation accuracy over standard ZNE. By integrating adaptive folding and selective filtering, our approach achieves more reliable and accurate noise extrapolation, even in circuits with high error variability. These findings suggest an improved method for ZNE that combines circuit-specific noise characterization with dynamic and selective mitigation techniques. We conclude that these innovations should become integral components of future error mitigation protocols for NISQ devices, ensuring more robust and scalable quantum computations. In this context, the term "robust" signifies the capacity of the proposed techniques to enhance the stability and efficacy of error reduction under variable noise conditions.

The article is organized as follows. After introducing known concepts and methods used in this article like standard ZNE (sZNE) and IC-ZNE in Sec.~\ref{sec: Concepts and Methods}, we introduce the noise adaptive folding and error scaling as well as the technique of noise filtering.
Finally, we demonstrate and explain the improved performance of our method compared to standard ZNE on IBMs quantum computing hardware in Sec. \ref{sec:results}.   
A conclusion is drawn in Sec. \ref{sec:conclusion}.

\section{\label{sec: Concepts and Methods}Concepts and Methods}

\subsection{\label{sec:sZNE}Zero-Noise Extrapolation}
Zero-noise extrapolation (ZNE) is a technique used to mitigate the effects of noise in quantum computations. It operates on the principle of leveraging multiple noisy circuit executions at varying noise levels to estimate the ideal, noise-free outcome. The core idea is to scale the strength of the noise in the quantum operations by implementing the circuit multiple times with different error rates.
To implement ZNE, one typically replaces the unitary operation $U$ by $U\cdot(U^\dagger\cdot U)^{n}$ \cite{He2020, Giurgica2020}, where $U^\dagger\cdot U$ is the identity. In this context, $U$ can be interpreted as the entire circuit, representing global folding, or alternatively, a specific region or gate, corresponding to local folding. As the dominant source of errors in the superconducting devices provided by IBM are the two-qubit gates \cite{Chow2011, Sheldon_2016, Kandala_2021, Magesan_2011,Magesan_2012}, the amplification is done by inserting additional CNOT gates.
By measuring the outputs of these altered circuits, one can construct a series of data points correlating the observed results with the corresponding amplification factor
\begin{equation}
    \lambda=2n+1,
    \label{eq:lambda}
\end{equation}
where $n$ denotes a non-negative integer, which is used to specify the number of the identity insertions. The standard scaling factors are $\lambda\in [1,3,5]$. Test runs with these scaling factors are referred to as sZNE, short for standard ZNE. An exponential fitting procedure as in \cite{Koenig_2024}, is then employed to extrapolate the expected result in the limit of zero noise. In this article, the discrete ZNE is used as a benchmark for further improvements in Sec.~\ref{sec: Concepts and Methods} C and D.

\subsection{\label{sec:IC-ZNE}Inverted-Circuit Zero-Noise Extrapolation}
The inverted-circuit ZNE (IC-ZNE)\cite{Koenig_2024} is a more refined version of the ZNE, where instead of assuming that the error is scaled by a factor $\lambda$ \cite{majumdar2023, Schultz2022}, the error strength $\epsilon$ is measured with an additional circuit, where the inverse of the circuit is added $U U^\dagger$ and the probability $P_0$ for every qubit to be in state $|0\rangle$ is measured. The noise strength can then be deduced with
\begin{equation}
    \epsilon=\frac{1-\sqrt{P_0-\frac{(1-P_0)}{2^q}}}{1+\frac{1}{2^q}}
    \label{eq:epsilon}
\end{equation}
where $q$ is the number of qubits. The expectation value $\langle A \rangle$ is then plotted as a function of $\epsilon$ and extrapolated to $\epsilon \to 0$ using a linear fit. 
In the following, we will use this method to measure the initial error strength $\epsilon_0$ for every unscaled circuit.

\subsection{\label{sec:ASF}Noise adaptive folding and error scaling}
\begin{figure}
    \includegraphics[width=0.5\textwidth]{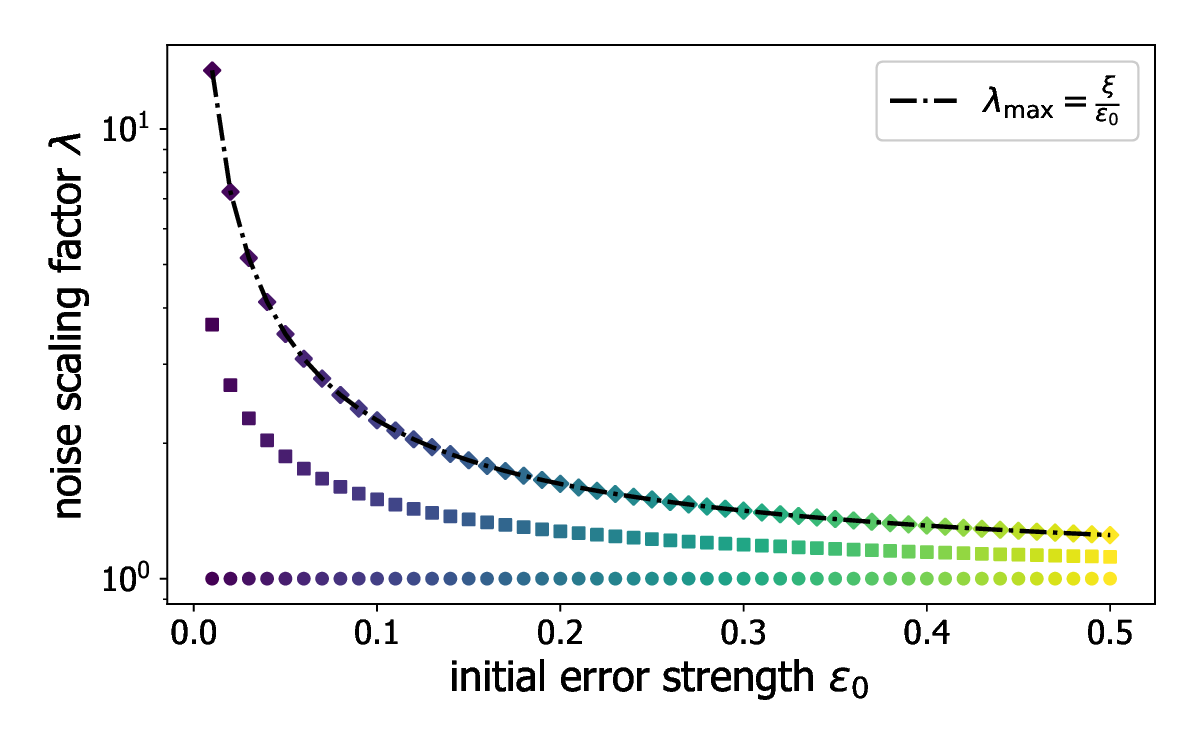}
    \caption{Scaling factors $\lambda_j$, $j=1,2,3$, as a function of the initial error strengths $\epsilon_0$ with exponential spacing. The black dash-dotted line indicates the maximum scaling factor $\lambda_{max}$, defining the upper limit of amplification achievable under the given conditions. Circles represent $\lambda_1=1$, squares for $\lambda_2$ and diamonds for $\lambda_3$.}
    \label{fig:asf vs epsilon 0}
\end{figure}
As outlined in Sec~\ref{sec:sZNE}, error scaling can be executed through the utilization of odd integers. More flexibility, however, can be achieved by realizing
arbitrary scaling factors that deviate from odd integers. This section presents corresponding
methodologies for amplifying errors in quantum circuits. These techniques facilitate the usage of, both, odd integer scaling factors and a random, non-uniform distribution of identity insertions for each CNOT gate \cite{He2020, Giurgica2020, Pascuzzi2022}.
For the latter purpose,
a more generalized equation for Eq.~\ref{eq:lambda}, can be written as
\begin{equation}
    \lambda=\frac{1}{N_c}\sum^{N_c}_{i=1}(2n_i+1),
    \label{eq:lambda_N}
\end{equation}
where $N_c$ is the number of CNOT gates in the original unscaled circuit and $n_i$ the number of identity insertions for the $i$-th CNOT gate.

It should be noted that not all scaling factors can be realized due to the finite number of CNOT gates in a circuit. For a given desired scaling factor $\lambda'$, we therefore first determine the closest scaling factor which can actually be reached according to Eq.~\ref{eq:lambda_N} by employing a rounding scheme that utilizes the floor function
\begin{equation}
    \lambda =\frac{\big\lfloor\frac{\lambda'-1}{2}N_c +\frac{1}{2}\big\rfloor}{N_c}+1.
    \label{eq:n_real}
\end{equation}
Then, we select a configuration of $n_i$'s according to Eq.~\ref{eq:lambda_N}. 
%tw Various choices are possible that all give rise to the same amplification factor. However, m
Multiple circuit configurations can be used that give rise to the same amplification factor, but with different CNOT gates
-- exhibiting potentially different gate errors --
being amplified.
In order to distribute the number of identity insertions across the various CNOT gates therefore as uniformly as possible, we choose $n_i \in \{\lfloor n\rfloor,\lfloor n\rfloor+1\}$, where $n=(\lambda-1)/2$, see Eq.~\ref{eq:lambda}. Thereby, the values of the $n_i$'s are fixed up to a permutation, which we select randomly. 

As an example, consider the case where the desired amplification factor is $\lambda'=4.2$ for $N_c=3$ CNOT gates. According to Eq.~\ref{eq:n_real}, the closest actual factor results as $\lambda=13/3$, giving rise to three possible configurations
\begin{equation}
	(n_1,n_2,n_3) \in \{(1,2,2),(2,1,2),(2,2,1)\}.
\end{equation}

Concerning the choice of desired scaling factors, we choose exponential scaling, as outlined in \cite{Krebs2022}, since this method assigns a greater weight to the regime of small amplification factors, where the extrapolation is expected to be more accurate. The exponential spacing factors can be calculated by
\begin{equation}
    \lambda'_j=(\lambda_\text{max})^{\frac{j-1}{k-1}},
    \label{eq:exp lambda}
\end{equation}
where $k$ is the number of amplification factors and $\lambda_\text{max}$ its maximum value.

In this paper, we propose an adaptation of the latter that incorporates a maximum noise scaling factor. This factor is based specifically on the measured or estimated error strength of the circuit, denoted as $\epsilon_0$. This aspect has not been explored in prior work. This adaptation addresses two critical issues. Firstly, excessive errors may amplify noise to the extent that the expectation value converges to that of a completely mixed state, rendering it unsuitable for extrapolation. Secondly, insufficient amplification can lead to data points that are too closely spaced, complicating the extrapolation process.
Consequently, a maximum noise scaling factor is defined dependent on the inverse of the initially measured or estimated error strength of the circuit $\epsilon_0$
\begin{equation}
    \lambda_\text{max}=\frac{\xi}{\epsilon_0},
    \label{eq:lambda_max}
\end{equation}
with $\xi=1/8$, determined based on the results obtained from experiments conducted with various parameter values. Results of the experiments can be found in Appendix~\ref{sec:xi}. The initial error strength $\epsilon_0$ can be measured as outlined in Sec.~\ref{sec:IC-ZNE} at the beginning of each ZNE cycle, without any amplification. The adaptive scaling factors (ASF) are then denoted as ASF B. Alternatively, $\epsilon_0$ can be estimated by the backend calibration data provided, where the errors of all CNOT gates occurring in the given circuit are added. The scaling factors thereby obtained are subsequently denoted by ASF M.

Figure~\ref{fig:asf vs epsilon 0} illustrates the relationship between ASF and the initial error strength $\epsilon_0$. It is observed that, as the initial error rate decreases, the range of scaling factors widens, indicating greater flexibility in amplification strategies. The data presented corresponds to a configuration with 
$N_c=100$ CNOT gates, where the difference between $\lambda$ and $\lambda'$ due to rounding, see Eq.~(\ref{eq:n_real}), is negligible.

The step-by-step procedure for applying noise-adaptive folding is as follows:
\begin{itemize}
    \item The maximum scaling factor, $\lambda_{\text{max}}$, is determined based on Eq.~\ref{eq:lambda_max}, either by estimating the initial error or measure it.
    \item Given the number of amplification factors, the exponential spacing is computed with Eq.~\ref{eq:exp lambda}.
    \item The actual amplification factors are adjusted according to the number of foldable CNOT gates in the circuit to maintain a physically realizable noise scaling according to Eq.~\ref{eq:n_real}.
    \item The various circuits are implemented on the quantum device, and expectation values are calculated and extrapolated to the zero-noise value.
\end{itemize}

\subsection{\label{sec:filtering}Noise filtering}
\begin{figure}
    \includegraphics[width=0.49\textwidth]{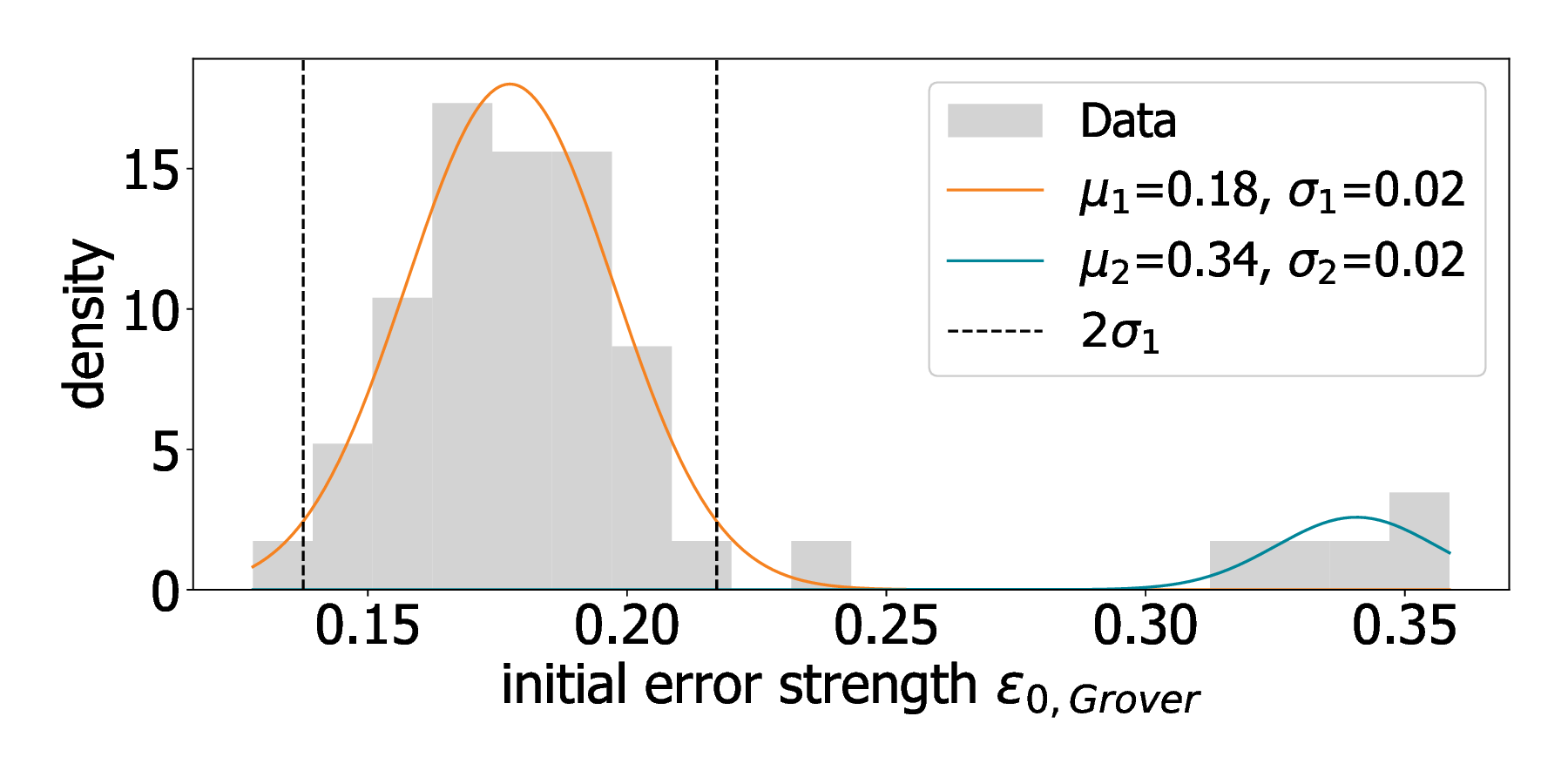}
    \caption{Distribution of 50 initial error strengths $\epsilon_0$ measured from the Grover circuit (see Appendix~\ref{sec:transpiled_circuits}). The data exhibit two Gaussian distributions, with the smaller distribution identified as an outlier and subject to removal through filtering. For the primary Gaussian distribution, values that exceed $2\sigma_1$ (vertical black dotted lines) from the mean are excluded from the dataset to enhance the reliability of the analysis.}
    \label{fig:Gaussian}
\end{figure}
While error mitigation techniques in quantum computing have made significant progress, the use of classical filtering methods to improve the reliability of mitigated results remains largely unexplored. 

Here, we demonstrate that the application of filtering enhances the precision of zero-noise extrapolation by identifying and discarding measurement results with disproportionately high error contributions. We implement a post-processing technique that evaluates circuit-level error rates or detects statistical outliers across repeated runs similar to \cite{Kim2023_util}.

However, our approach predominantly focuses on the initial error strength, which has not been explored in previous work. By directly measuring the initial error strength, $\epsilon_0$, of a circuit—yielding a set of baseline values $\epsilon_0$—we can systematically reject unreliable experiments. Under ideal conditions, these $\epsilon_0$ values are expected to follow a normal Gaussian distribution
\begin{equation}
    p(\epsilon_0)=\frac{1}{\sqrt{2\pi\sigma}}\exp\Bigg(- \frac{(\epsilon_0-\mu)^2}{2\sigma^2}\Bigg),
\end{equation}
where $\mu$ is the mean error and $\sigma$ is the standard deviation.

In actual environments, deviations from this ideal behavior, such as outliers or the emergence of a secondary distribution, may arise due to rare noise events or transient hardware instabilities. 
Notably, we introduce the assumption of a secondary Gaussian distribution in our analysis, which has not been previously considered.
An example is shown in Fig.~\ref{fig:Gaussian}, where 50 independent ZNE runs of the Grover circuit (represented in Sec.~\ref{sec:circuits}) are analyzed. In each run, the initial error strength is measured prior to noise amplification. Runs associated with a secondary distribution or strong deviations from the main Gaussian are excluded from further analysis.

To identify potential secondary error contributions in the distribution of initial error strengths, we apply a Gaussian Mixture Model (GMM) \cite{Bishop2006}. A GMM is a probabilistic model that assumes the data is drawn from a weighted sum of multiple Gaussian distributions. Formally, the probability density function of a GMM with $T$ components is defined as
\begin{equation}
    p(x)=\sum_{t=1}^T\pi_t\cdot \mathcal{N}(x|\mu_t, \sigma^2_t),
\end{equation}
where $\pi_t$ denotes the mixing coefficient of the $t$-th component (with $\sum_{t=1}^T\pi_t=1$), and $\mathcal{N}(x|\mu_t, \sigma^2_t)$ is the Gaussian distribution with mean $\mu_t$ and the variance $\sigma^2_t$. The parameters $\{\mu_t, \pi_t,\sigma_t\}$ are estimated via the Expectation-Maximization (EM) algorithm, which iteratively refines the assignment of data points to components based on maximum likelihood estimation. We use the implementation provided by the \textit{Scikit-learn} Python library \cite{Pedregosa2011}.

In our analysis, we fit a two-component GMM ($T=2$) to the distribution of measured initial error strengths, aggregated across 50 independent repetitions of ZNE runs.
To identify outliers that deviate too strongly from the primary component,
entire ZNE runs are excluded from further analysis if the measured initial error strength $\epsilon_0$ satisfies the condition
\begin{equation}
    |\epsilon_0 - \mu_1| > 2\sigma_1,
\end{equation}
where $\mu_1$ and $\sigma_1$ denote the mean and standard deviation of the primary Gaussian component. This additional step helps mitigate the influence of ambiguous points.

The entire run is excluded from further analysis, including all measurements across amplification factors and twirling realizations associated with that run. This dual-filtering approach ensures that only consistent and statistically reliable data contribute to the extrapolation, thereby improving the robustness and accuracy of the ZNE method. 

We compare the $\epsilon_0$-based filtering strategy to an alternative global filtering approach applied directly to the expectation values obtained in the error mitigation experiments. In this method, the distribution of measured expectation values $\langle A \rangle$ is analyzed separately for each noise scaling factor $\lambda_i$. Specifically, across all 50 experimental repetitions, the set of expectation values for each $\lambda_i$ is assumed,
under ideal conditions,
to follow a normal distribution centered around a mean value $\mu_A$ with standard deviation $\sigma_A$. The resulting Gaussian distribution includes data from all twirling instances across all runs, thereby capturing the statistical spread of outcomes for that particular noise level. Additionally, the implementation of global filtering involves the removal of data points associated with a secondary Gaussian by utilizing a GMM. 

Individual measurements that deviate significantly from the mean are then identified as statistical outliers. In particular, any data point satisfying the condition
\begin{equation}
    |\langle A \rangle - \mu_A| > 2\sigma_A,
\end{equation}
(where $\mu_A$ and $\sigma_A$ again refer to the primary Gaussian component)
is considered to be potentially affected by sporadic error events, such as rare hardware fluctuations, transient gate errors, or readout anomalies, and is excluded from the extrapolation procedure. In contrast to $\epsilon_0$-based filtering, which removes entire runs based on the baseline error strength, this global filtering approach selectively discards only anomalous individual data points at each noise level.

The two filtering techniques ensure that the extrapolation process is based solely on data that accurately reflects the intrinsic noise characteristics of the circuit. As demonstrated in Fig.~\ref{fig:Gaussian}, this filtering eliminates outlier contributions that could otherwise distort the final zero-noise estimate.

\subsection{\label{sec:circuits}Sample circuits and test parameters}
This section provides a detailed description of the three distinct circuits executed on quantum hardware, as well as the techniques that are utilized in the execution of said circuits.
The first circuit is an example of Grover's quantum search algorithm \cite{grover1996} with three qubits and 10 CNOT gates after the transpilation process to the considered device, here \textit{ibmq\_ehningen}. The device properties can be found in Appendix~\ref{sec:properties}. The transpiled circuit is shown in Appendix~\ref{sec:transpiled_circuits}). 
We consider the observable 
\begin{equation}
    A_{\text{Grover}}=|101\rangle \langle 101|+|011\rangle \langle 011|
    \label{eq:ew_grover}
\end{equation}
corresponding to the probability of measuring one of the two solutions 101 or 011 of the search problem encoded by the oracle of the circuit. Without errors, we therefore obtain $\langle A_{\text{Grover}}\rangle_{\text{ideal}}=1$.

The second circuit is taken from the Harrow-Hassidim-Lloyd algorithm (HHL algorithm) \cite{Harr2009}. This algorithm can be used to solve linear systems of equations defined by a matrix $B\in\mathbb{C}^{N\times N}$ and a vector $\vec{b}\in\mathbb{C}^{N}$ to find $\vec{x}\in\mathbb{C}^{N}$, so that $B\Vec{x}=\Vec{b}$. We use an example with $N=2$ as described in \cite{HHL-IBM} and
\begin{equation}
    B=\begin{pmatrix}1 & -1/3\\-1/3 & 1 \end{pmatrix},\quad 
    \vec{b}=\begin{pmatrix}1 \\ 0\end{pmatrix}.
\end{equation}
The corresponding transpiled circuit for four qubits exhibits 18 CNOT gates and can be found in Appendix~\ref{sec:transpiled_circuits}. Only the last qubit is measured, and the corresponding expectation value of the observable 
\begin{equation}
        A_\text{HHL} =\mathbb{I} \otimes \mathbb{I} \otimes \mathbb{I} \otimes |1\rangle\langle 1|
        \label{eq:ew_hhl}
\end{equation}
i.e., the probability to measure the last qubit in state $|1\rangle$,
yields the norm of $\vec{x}$ 
as follows:
\begin{equation}
    ||\vec{x}||=\frac{3}{2}\sqrt{\langle A_{\text{HHL}}  \rangle_{\text{ideal}}}.
\end{equation}
In our case, $\langle A_{\text{HHL}}  \rangle_{\text{ideal}}=5/8$, in accordance with the correct solution $\vec{x}=(9/8,3/8)$ of the above linear system.

The third circuit is the Hadamard-Ladder quantum circuit (H-Ladder) as used in \cite{ketterer2023, Brandhofer2023}, which consists of eight qubits and a number of single-qubit gates and seven CNOT gates. The corresponding transpiled circuit can also be found in the Appendix~\ref{sec:transpiled_circuits}. The considered expectation value is
\begin{equation}
    \langle A_{\text{Ladder}}\rangle=\langle 11111111|\rho_1|11111111\rangle
\end{equation}
with  $\langle A_{\text{Ladder}}\rangle_\text{ideal}=0.5$.

In all cases (unless otherwise specified), we use the following parameters: the original circuits correspond to $\lambda=1$. We generate two scaled circuits with scaling factors $\lambda=3$ and 5 (for sZNE) by multiplying CNOT gates, as explained in Sec.~\ref{sec:sZNE}. For adaptive scaling factors, the maximum scaling factor is first determined by the initial error strength, after which the scaling factors are selected from the range $\lambda \in [1,~\lambda_{max}]$ as detailed in Sec.~\ref{sec:ASF}, encompassing three distinct factors. Randomized compiling is used always, as described in \cite{Koenig_2024, Wall2016, Gottesman1998, Zhao2022, Rudinger2021, ketterer2023}. 16 different twirled versions of each circuit are randomly generated and executed with $10~000/16=625$ shots, from which the expectation values $\langle A\rangle$ are determined separately for each of the 16 twirled versions. For circuits with scaling factor $\lambda>1$, different random permutations of the $n_i$'s are chosen for each twirled version, as explained in Sec.~\ref{sec:ASF}.

In the case of a $\epsilon_0$-measurement, the initial circuit with its inverse is run once with $10~000$ shots. Thereby, we obtain in total $3\times 16=48$ data points $(\lambda,\langle A\rangle)$, which are fitted by an exponential function to obtain the extrapolated noise-free value of $\langle A\rangle$ at $\lambda=0$, respectively. In addition, read-out error mitigation is applied based on the method M3 \cite{Nation2021}.

\section{\label{sec:results}Performance Analysis of Adaptive Scaling and Filtering methods}
\begin{figure}
    \includegraphics[width=0.5\textwidth]{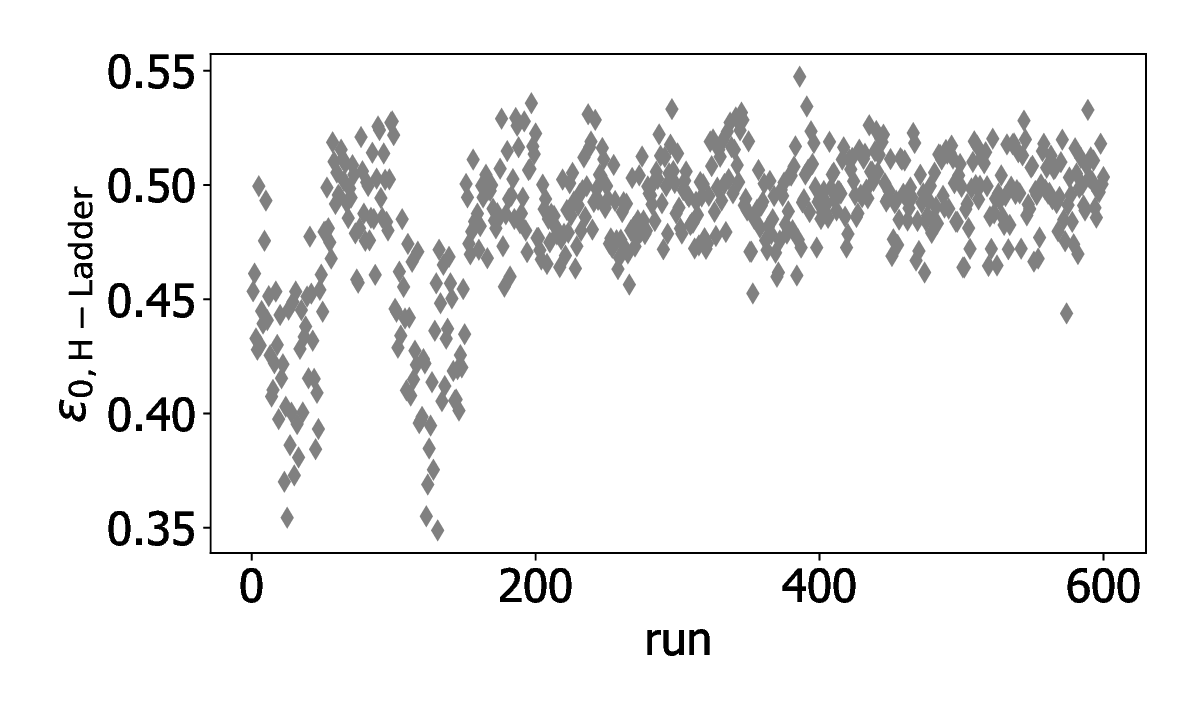}
    \caption{Variation in error strength across multiple backend executions for the Hadamard-Ladder algorithm. The plot depicts the measured error strength over successive runs on the same backend, revealing substantial fluctuations in error rates between runs. Notably, the Ladder algorithm exhibits consistently higher error rates beyond the 150th run. These variations underscore the necessity of accounting for dynamic error characteristics in quantum algorithm execution and error mitigation strategies.}
    \label{fig:epsilon}
\end{figure}

The inverted-circuit approach allows for direct measurement of the initial error strength $\epsilon_0$ without artificially amplifying errors, as outlined in Sec.~\ref{sec:IC-ZNE}. Figure~\ref{fig:epsilon} illustrates that the initial error exhibits substantial fluctuations between experimental runs for the H-Ladder and displays significant discontinuities before the 150th run. These variations likely stem from backend-induced errors, including both non-unitary and coherent error contributions. Consequently, it is essential to adjust the scaling or measure the error strength for each scaling factor, as emphasized in Ref.~\cite{Koenig_2024}.

\begin{figure}
    \includegraphics[width=0.48\textwidth]{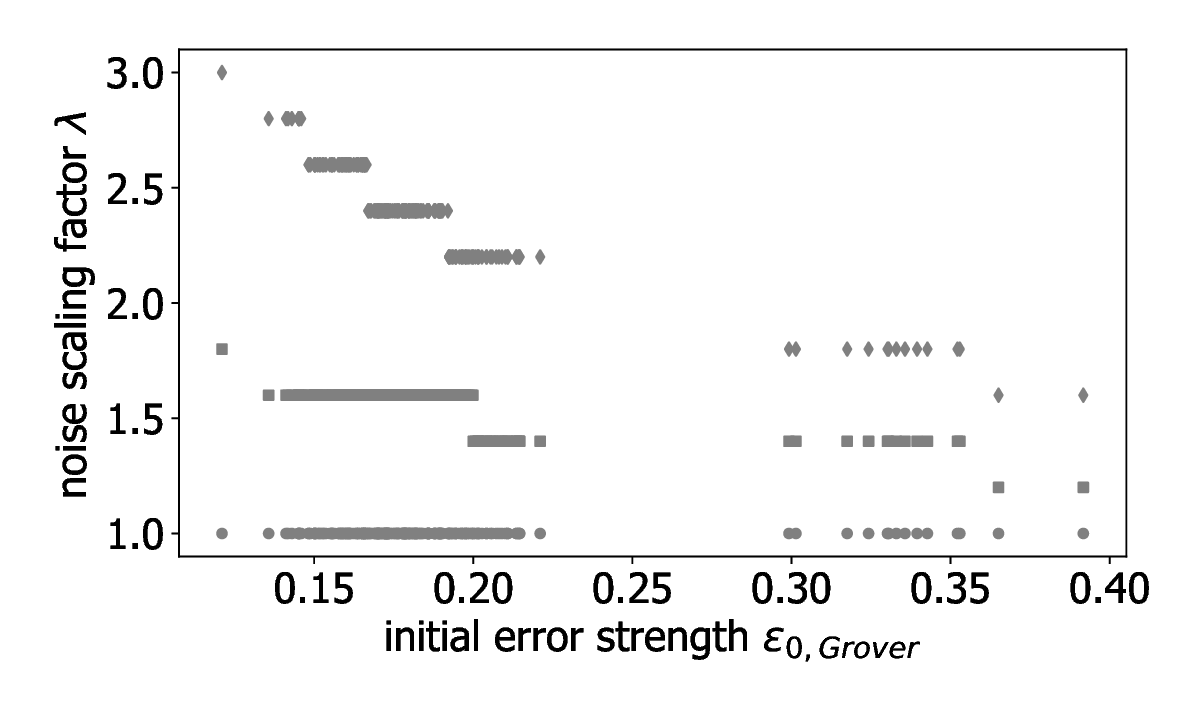}
    \caption{Noise scaling factors for the Grover algorithm of 150 different runs. The real factors are dependent on the number of CNOT gates, here $N_c=10$. A higher initial error strength $\epsilon_0$ leads to a lower maximum noise scaling factor $\lambda_{\text{max}}$. Circles represent $\lambda_1 = 1$, squares for $\lambda_2$ and diamonds for $\lambda_3$.}
    \label{fig:asf vs eps device}
\end{figure}

\begin{figure*}
    \includegraphics[width=\textwidth]{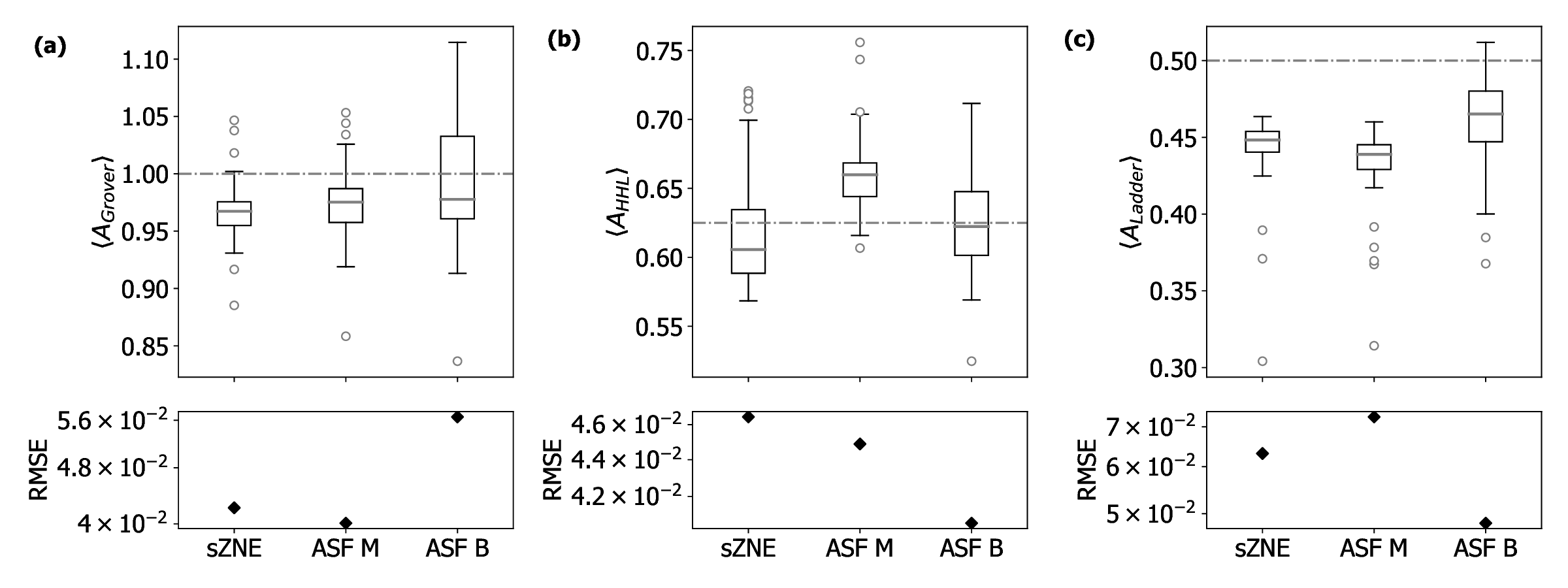}
    \caption{Comparison of the performance of standard ZNE for the (a) Grover, (b) HHL and (c) H-Ladder algorithms with adaptive scaling factors (ASF). The exact result is indicated by the dotted horizontal line. The lower subplots depict the root mean square errors (RMSE). Each subplot presents results for sZNE and ASF where $\epsilon_0$ is either estimated from calibration data (ASF M) or measured with an inverted circuit (ASF B). Overall, ASF methods achieve lower RMSE values across most circuits. However, in the case of Grover using measured $\epsilon_0$ leads to a degradation in performance, and a similar trend is observed for the H-Ladder algorithm when using estimated $\epsilon_0$.}
    \label{fig:asf backend}
\end{figure*}

\begin{figure}
    \includegraphics[width=0.5\textwidth]{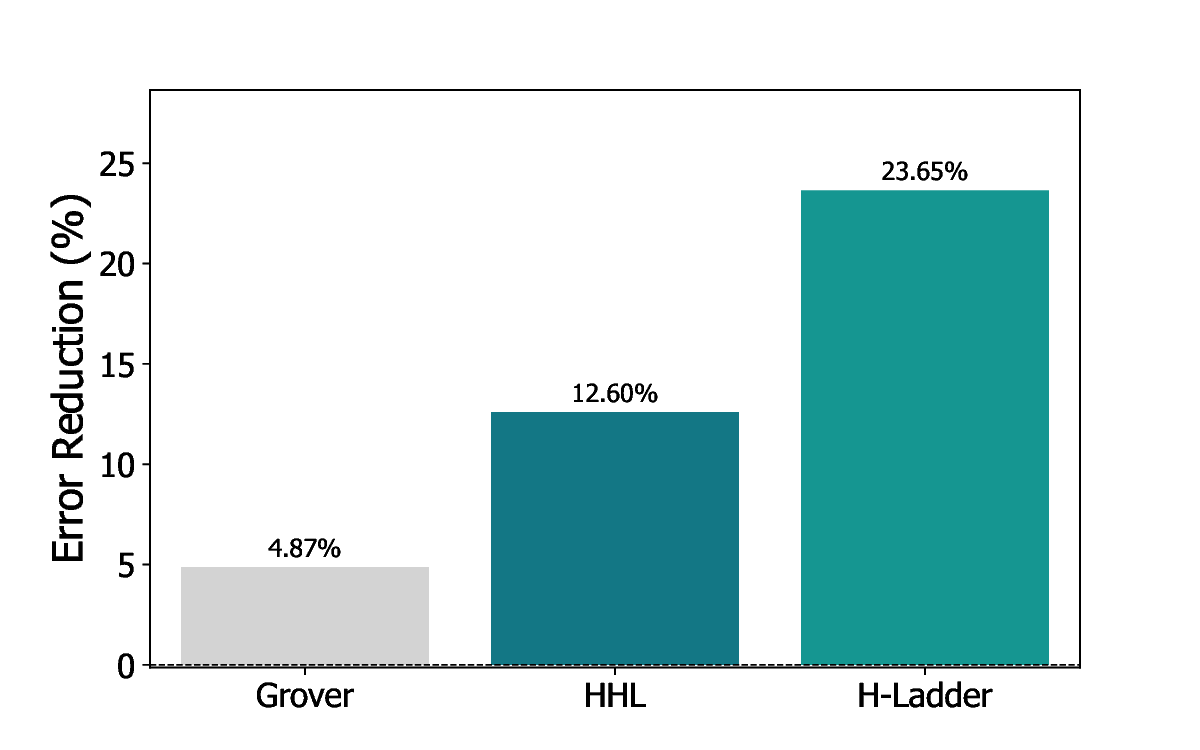}
    \caption{Relative error reduction achieved using adaptive scaling factors (ASF) for the Grover, HHL, and H-Ladder algorithms, as compared to standard ZNE (based on the data shown in Fig.~\ref{fig:asf backend}),
    where ASF M is employed for Grover and ASF B for HHL and H-Ladder. 
    The most significant improvement overall is achieved for the H-Ladder algorithm, resulting in a 23.65\% decrease in RMSE.}
    \label{fig:error reduction asf}
\end{figure}

To address this challenge, the dynamic nature of these errors is incorporated through adaptive scaling factors, as detailed in Sec.~\ref{sec: Concepts and Methods}. Specifically, the scaling factors are adjusted exponentially based on the initial error strength, $\epsilon_0$. This approach is demonstrated in Fig.~\ref{fig:asf vs eps device} for the Grover circuit, where initial error strengths range from 0.1 to 0.4, and the corresponding adaptive scaling factors vary from 1 to 3. Notably, when error strengths are inferred from backend-reported error data, the scaling factors tend to be larger due to optimistic assumptions about lower error rates. In contrast, direct measurements of error strengths yield lower scaling factors, reflecting the presence of higher actual error rates.

The results of $M=50$ runs on IBM's quantum device $ibmq\_ehningen$ per algorithm are shown  in Fig.~\ref{fig:asf backend}.
In each case, we obtain 50 extrapolated values $\langle A_{\text{Grover}}\rangle_i$, $\langle A_{\text{HHL}}\rangle_i$ and $\langle A_{\text{Ladder}\rangle_i}$ from the different runs ($i=1,2,\dots,M)$. From these values, a box plot is generated showing the median value (central line) and the interquartile range (box). The whiskers extend to the highest and lowest values within 1.5 times the interquartile range. Any data points outside of these boundaries are considered outliers and are plotted separately as circles. The accompanying plots at the bottom display the root mean square deviation (RMSE) from the exact value:
\begin{equation}
    \text{RMSE}=\sqrt{\frac{1}{M}\sum_{i=1}^M \Bigl(\langle A\rangle_i-\langle A\rangle_{\text{ideal}}\Bigr)^2},
\end{equation}
where $A=A_{\text{Grover}}$, $A_{\text{HHL}}$ or $A_{\text{Ladder}}$, respectively. 
Each algorithm is benchmarked against standard zero-noise extrapolation (sZNE), with results obtained using adaptive scaling factors (ASF) derived either from calibration-based error estimates (ASF M) or from direct backend measurements (ASF B). In general, the use of ASF yields lower RMSE values across most circuits, indicating improved noise mitigation. However, notable exceptions are observed: for the Grover algorithm, ASF B leads to degraded performance, while for the H-Ladder algorithm, ASF M shows a similar decline. These deviations suggest that the effectiveness of the ASF approach depends on the accuracy of the initial error strength $\epsilon_0$. For algorithms with higher gate counts and consequently larger initial errors, such as HHL and H-Ladder, ASF B provides a more realistic characterization of noise. In contrast, for the Grover algorithm, which exhibits relatively low error rates, model-based estimates (ASF M) appear more reliable. Although ASF B tends to produce extrapolations closer to the ideal expectation value, the broader variance and presence of outliers result in a less favorable median outcome, ultimately leading to an increased RMSE.

For completeness, the results of using adaptive scaling factors for inverted-circuit ZNE (IC-ZNE) \cite{Koenig_2024} are shown in Appendix~\ref{sec:asf+iczne}. Here, we do not observe an improvement in performance, which can be explained by the fact that IC-ZNE is better able to deal with different levels of error.
In contrast, the effectiveness of ASF more strongly depends on the underlying noise characteristics and the structure of the quantum circuit.

Figure~\ref{fig:error reduction asf} illustrates the performance improvement achieved by applying ASF to sZNE, based on the data presented in Fig.~\ref{fig:asf backend}. The relative error reduction is computed by comparing the RMSE values obtained with ASF to that of sZNE as a baseline. The most significant improvement is observed for the H-Ladder algorithm using backend-based error measurement (ASF B), resulting in a 23.65\% reduction in RMSE. For the HHL algorithm, ASF B also yields the largest gain, while for Grover, the best result is achieved with model-based error estimates (ASF M), even though with a more modest improvement of 4.87\%. This smaller gain can be associated to the inherently lower error rates of the Grover circuit when using standard ZNE.

\begin{figure}
    \centering
    {\includegraphics[width=0.49\textwidth]{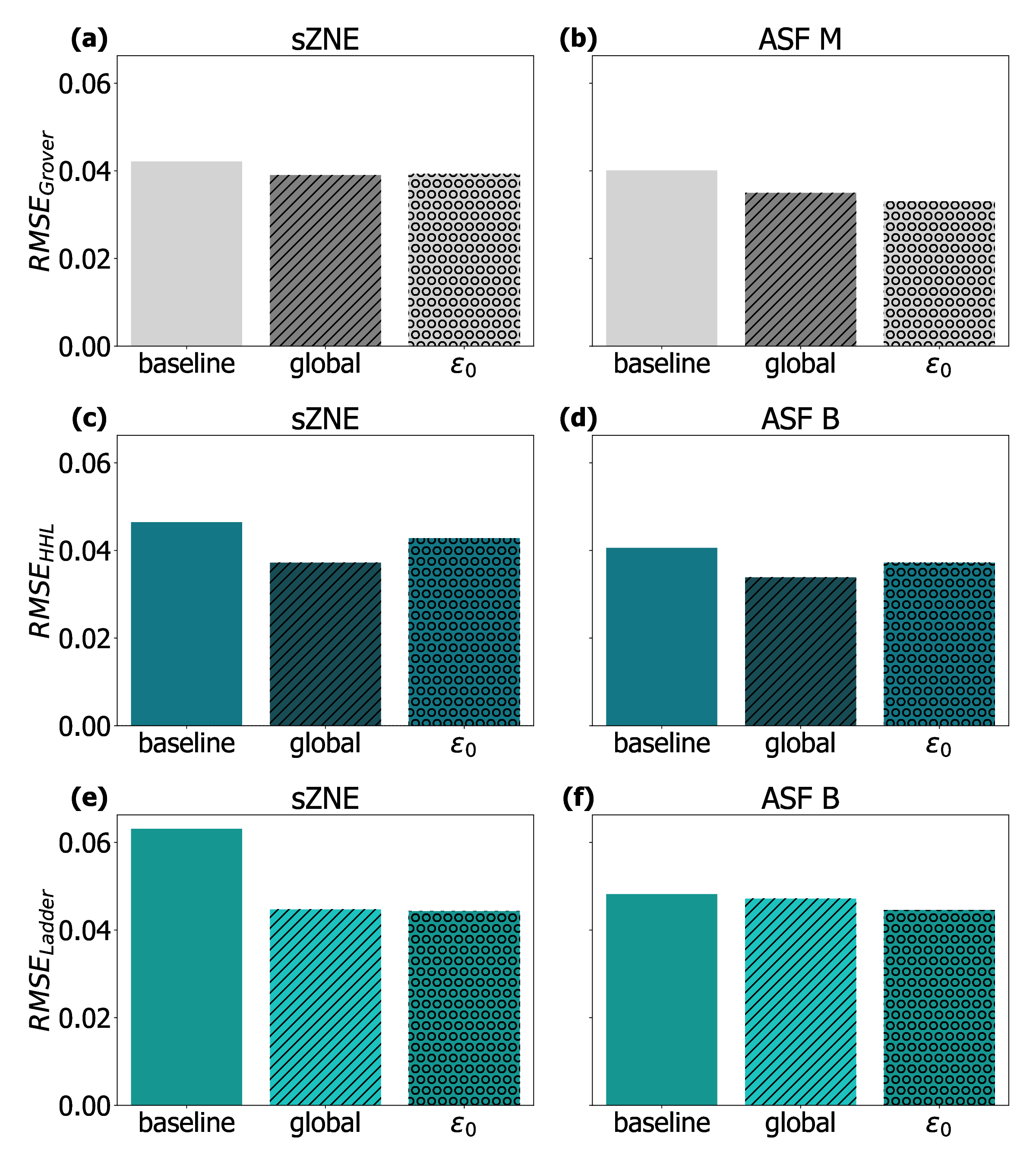}}
    \caption{RMSE by filtering the data either by global 
    (middle columns) or $\epsilon_0$ filtering (right columns), for the Grover algorithm (a,b), the HHL algorithm (c,d) and the H-Ladder (e,f), using standard ZNE (a,c,e) or adaptive scaling factors (b,d,f). As in Fig.~\ref{fig:error reduction asf}, the variant ASF-M (error strength $\epsilon_0$ estimated from calibration data) is chosen for Grover (b), and ASF-B ($\epsilon_0$ measured directly on backend) for the other two (d,f). To enable a direct comparison, the values without filtering (see Fig.~\ref{fig:asf backend}) are shown again (left columns).}
    \label{fig:RMSE filter}
\end{figure}

To evaluate the impact of the filtering techniques described in Sec.~\ref{sec:filtering}, we evaluate their effect on the RMSE values in zero-noise extrapolation, both with and without adaptive scaling factors. The objective is to improve the accuracy of the extrapolated results by systematically discarding outlier data.

Figure~\ref{fig:RMSE filter} presents a comparison of both filtering approaches and sZNE or ASF without filtering as baseline, allowing identification of the more effective method in each scenario. Global filtering reduces the influence of anomalous fluctuations by acting directly on the expectation values. In contrast, $\epsilon_0$ filtering discards data based on deviations in the initial error strength.

The implementation of filtering consistently improves the quality of zero-noise extrapolation with and without ASF across all evaluated circuits. The selection of a suitable filtering method, however, is a non-trivial task. In certain cases, global filtering provides superior results, while in others, $\epsilon_0$-based filtering is more effective.
For Grover, the best results are obtained using ASF M in combination with $\epsilon_0$ filtering; for HHL, ASF B with global filtering performs best; and for the Ladder circuit, $\epsilon_0$ filtering in conjunction with ASF yields the the lowest RMSE values.

The relative error reductions achieved by the two filtering methods are summarized in Tab~\ref{tab:error_reduction}, based on the RMSE values shown in Fig.~\ref{fig:RMSE filter}. The error reduction, including that for the ASF baseline, is expressed relative to the sZNE baseline. Within the table, the baseline for the ASF method is denoted by '-' in the filtering column. For Grover’s algorithm, global filtering achieves a higher reduction when ASF is used. This is likely due to imprecise error estimates used in ASF. Filtering can mitigate these inaccuracies. The largest improvement overall for Grover is achieved using $\epsilon_0$ filtering, with an error reduction of 21.68\%.

For the HHL algorithm, the optimal overall performance is achieved by ASF B with global filtering, resulting in a 27.12\% error reduction compared to the sZNE baseline value. Additionally, for the sZNE method, global filtering outperforms $\epsilon_0$ filtering, yielding a reduction of 19.89\% versus 7.86\%, respectively. This difference can be attributed to the presence of more prominent outliers in the global expectation values, while the directly measured initial error strength demonstrates lower variance.

In the case of the H-Ladder circuit, an enhancement is primarily seen with sZNE utilizing $\epsilon_0$ filtering, reflecting the circuit’s high initial error variance and sensitivity to noise fluctuations. This method demonstrates the highest overall improvement, achieving a reduction of 29.81\%. Adaptive scaling factors have been shown to enhance the baseline RMSE value. When utilized with the $\epsilon_0$ filtering method, this enhancement can contribute to the elimination of outliers. However, the efficacy of this approach is comparatively lower than that of the baseline sZNE and $\epsilon_0$ filtering method.

The use of IC-ZNE with the different filtering techniques does not improve its already good RMSE values significantly, since it already takes into account the different error levels and therefore has fewer outliers. Results for filtering techniques with IC-ZNE can be found in Appendix~\ref{sec:asf+iczne}.

These results underline the importance of incorporating filtering into quantum error mitigation pipelines. By removing inconsistent or extreme data points, both filtering strategies contribute to reduced RMSE values and improved precision of extrapolated values. Moreover, the synergy between ASF and filtering highlights the role of adaptive techniques in tailoring error mitigation to circuit-specific noise characteristics, thereby enhancing robustness and reliability of ZNE methods.

\begin{table}[h]
\caption{\label{tab:error_reduction}
Relative error reduction due to filtering for different algorithms and configurations. The baseline error reduction for the ASF method is denoted by a '--' in the filtering column.}
    \centering
    \begin{tabular*}{0.48 \textwidth}{@{\extracolsep{\fill}}cccc}
        \hline
        Algorithm & Method & Filtering & Error Reduction (\%) \\
         \hline
        \multirow{6}{*}{Grover} & \multirow{2}{*}{sZNE} & global       & 7.35   \\
                                &                       & $\epsilon_0$ & 6.76   \\
                                \cline{2-4}
                                &                       & -- & 4.87   \\
                                & ASF M                 & global       & 17.04  \\
                                &                       & $\epsilon_0$ & 21.68  \\
        \hline
        \multirow{6}{*}{HHL}    & \multirow{2}{*}{sZNE} & global       & 19.89  \\
                                &                       & $\epsilon_0$ & 7.86   \\
                                \cline{2-4}
                                &                       & -- & 12.60  \\
                                & ASF B                 & global       & 27.12  \\
                                &                       & $\epsilon_0$ & 19.90   \\
        \hline
        \multirow{6}{*}{H-Ladder} & \multirow{2}{*}{sZNE} & global       & 29.19  \\
                                &                       & $\epsilon_0$ & 29.81  \\
                                \cline{2-4}
                                &                       & -- & 23.65  \\
                                & ASF B                 & global       & 25.29   \\
                                &                       & $\epsilon_0$ & 29.38   \\
        \hline
    \end{tabular*}
\end{table}

\section{\label{sec:conclusion}Conclusions and Outlook}
In this work, we explored strategies to improve the accuracy and robustness of zero-noise extrapolation, a key technique in quantum error mitigation. We focused on two complementary approaches: the use of adaptive scaling factors that account for circuit-specific noise characteristics, and post-processing-based filtering methods to discard statistically inconsistent data. These strategies were benchmarked across the Grover, HHL, and H-Ladder algorithms.

Our results demonstrate that tailoring the extrapolation process to the noise profile of each circuit, either through error estimation from calibration data or through direct measurements using the inverted circuit method, can significantly reduce the root mean square error of mitigated results. The benefits of adaptive scaling are particularly pronounced for algorithms with higher CNOT counts and correspondingly greater noise levels, such as HHL and H-Ladder. However, we also observe that the effectiveness of the scaling strategy depends on both the circuit and the accuracy of the underlying noise estimates.

In addition, we evaluated two filtering techniques designed to eliminate unreliable data: global filtering based on fluctuations in expectation values, and $\epsilon_0$-based filtering that uses circuit-level error measurements. Both methods proved beneficial, with the optimal strategy depending on the algorithm and the specific error landscape. While global filtering can be applied purely in post-processing, measuring $\epsilon_0$ requires additional quantum resources. Nevertheless, both techniques significantly enhance the quality of extrapolated values, yielding RMSE reductions, when combined with adaptive scaling factors of up to 30\%.

Looking ahead, these results open up promising directions for further refinement of error mitigation methods. Future work could extend these techniques to a wider set of algorithms, noise models, and quantum hardware platforms. More broadly, our findings highlight the value of dynamically adapting error mitigation to circuit-specific behavior, and they establish a framework for integrating classical data analysis with quantum experiments to push the limits of achievable precision on near-term quantum devices.

\hfill\break
\hfill\break
The data that support the findings of this study are available from the corresponding author, KK, upon reasonable request.

\begin{acknowledgments}
The authors acknowledge funding from the Ministry of Economic Affairs, Labour and Tourism Baden-W{\"u}rttemberg, under the project Sequoia - End to End in the frame of the Competence Center Quantum Computing Baden-W{\"u}rttemberg.
\end{acknowledgments}

%\nocite{*}
\typeout{}
\clearpage
\bibliography{main}
\clearpage

\appendix
\section{\label{sec:xi}Choice of maximum scaling factor}

\begin{figure}[ht]
    \centering
    \includegraphics[width=\linewidth]{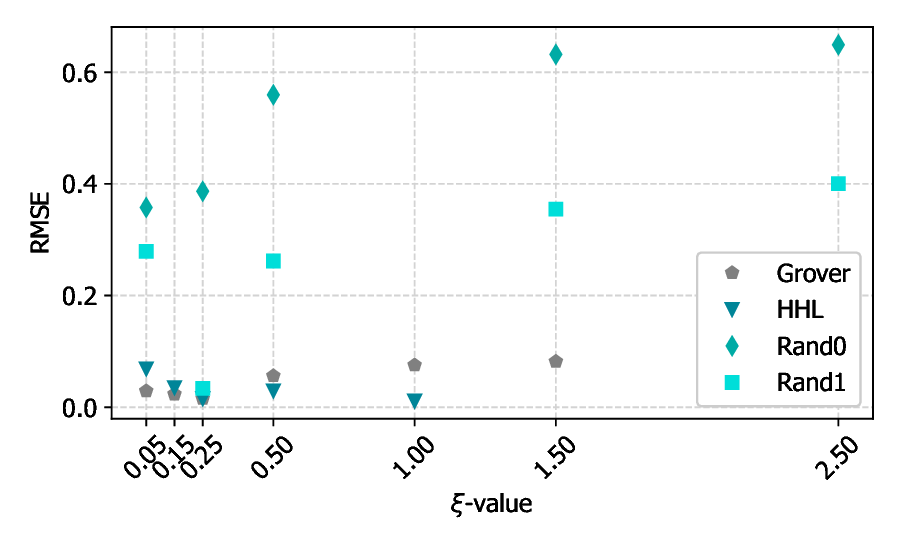}
    \caption{Influence of different $\xi$-values for Grover and HHL circuits, as well as for two randomly generated circuits with four qubits and eight CNOT gates. Optimal values for the parameter $\xi$ range from 0.05 to 0.25 to yield the lowest RMSE values.}
    \label{fig:xi}
\end{figure}

In Fig. \ref{fig:xi} different values for $\xi$ are investigated for Grover, HHL and two random circuits. The Grover and HHL circuits are 
%tw displayed on 
described in
Sec.~\ref{sec:circuits}. The two random circuits under consideration consist of four qubits and eight CNOT gates. The findings of these test runs suggest that optimal values for the parameter $\xi$ range from 0.05 to 0.25, in terms of achieving minimal RMSE values for all considered circuits.

\section{\label{sec:properties}Device properties}
\begin{figure}
    \includegraphics[width=0.45\textwidth]{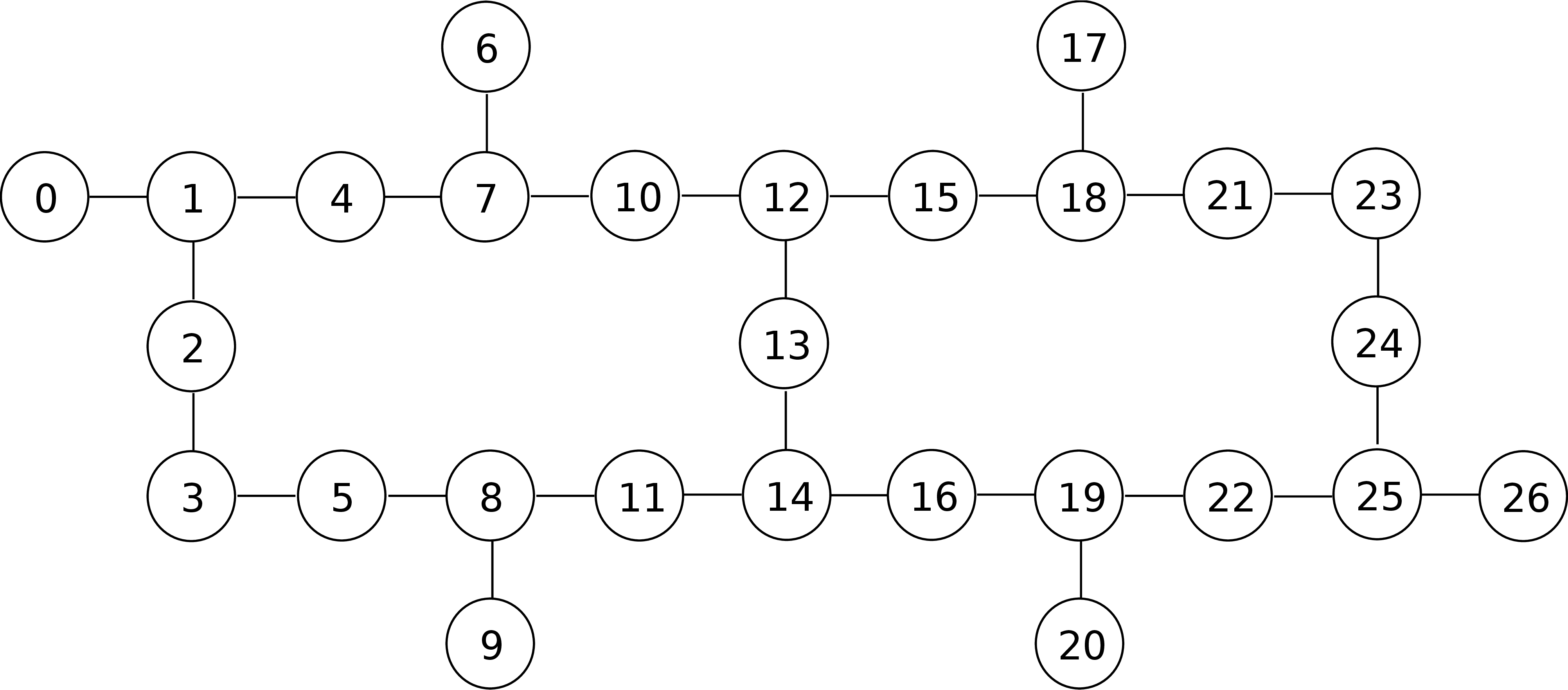}
    \caption{Connectivity (black lines) between 27 qubits (black circles) of IBM's Ehningen device.}
    \label{fig:layout}
\end{figure}

This section presents the detailed specifications of the data obtained from IBM's quantum device, as described in the main text. Fig.~\ref{fig:layout} illustrates the connectivity of the device \textit{ibmq\_ehningen}, which hosts 27 qubits in a heavy hexagonal lattice.

The error rates of all CNOT gates of \textit{ibmq\_ehningen} (from August 18th, 2023) as reported by IBM are displayed in Table~\ref{table:cxgate_properties}. We observe variations in error rates across different CNOT gates, which can significantly impact the success probability of quantum algorithms. The median error rate is 0.705\%. The CNOT gates used for runs involving the HHL circuit are ([4, 1],[6, 7], [7, 6],[4, 7],[7, 4],[4, 1],[6, 7]), while those employed for runs with the Grover circuit are ([4, 7],[6, 7],[7, 6]) and the CNOT gates of the H-Ladder ([12,10],[13,12],[14,13],[14,16],[16,19],[19,22],[22,25]).

Note that these error rates only give a partial description of the noise occurring on the real device (e.g., they do not include crosstalk effects \cite{ketterer2023}). Therefore, the total error strengths measured using the method of inverted circuits ($\epsilon_0\simeq 0.15$ for the Grover circuit and $\epsilon_0\simeq 0.4$ for the HHL circuit are typically larger than expected from the error rates per CNOT gate.  

The single-qubit device properties of \textit{ibmq\_ehningen} as reported by IBM are presented in Table~\ref{table:backend_properties}.

\begin{table}
    \centering
    \caption{Table of the gate error of the available two-qubit gate pairs for \textit{ibmq\_ehningen} (as of August 18th, 2023).}
    \label{table:cxgate_properties}
    \begin{tabular*}{0.5 \textwidth}{@{\extracolsep{\fill}}cccc}
        CX Gate Pair & Gate error & CX Gate Pair & Gate error \\ \hline
        25\_22 & 0.00565 & 10\_12 & 0.00849 \\ 
        22\_25 & 0.00565 & 12\_10 & 0.00849 \\ 
        20\_19 & 0.00602 & 7\_10 & 0.02103 \\
        19\_20 & 0.00602 & 10\_7 & 0.02103 \\ 
        14\_16 & 0.00847 & 19\_22 & 0.00993 \\ 
        16\_14 & 0.00847 & 22\_19 & 0.00993 \\ 
        18\_17 & 0.00896 & 21\_23 & 0.00961 \\ 
        17\_18 & 0.00896 & 23\_21 & 0.00961 \\ 
        14\_11 & 0.00959 & 13\_12 & 0.00629 \\ 
        11\_14 & 0.00959 & 12\_13 & 0.00629 \\ 
        12\_15 & 0.00726 & 26\_25 & 0.00684 \\ 
        15\_12 & 0.00726 & 25\_26 & 0.00684 \\
        8\_5 & 0.0101 & 3\_2 & 0.00358 \\ 
        5\_8 & 0.0101 & 2\_3 & 0.00358 \\ 
        1\_0 & 0.01138 & 7\_6 & 0.00595 \\ 
        0\_1 & 0.01138 & 6\_7 & 0.00595 \\ 
        1\_2 & 0.00681 & 16\_19 & 0.00676 \\
        2\_1 & 0.00681 & 19\_16 & 0.00676 \\ 
        8\_11 & 0.01272 & 18\_21 & 0.00607 \\
        11\_8 & 0.01272 & 21\_18 & 0.00607 \\
        24\_23 & 0.00885 & 4\_7 & 0.00542 \\ 
        23\_24 & 0.00885 & 7\_4 & 0.00542 \\ 
        4\_1 & 0.00655 & 14\_13 & 0.00667 \\ 
        1\_4 & 0.00655 & 13\_14 & 0.00667 \\ 
        8\_9 & 0.03875 & 24\_25 & 0.00839 \\ 
        9\_8 & 0.03875 & 25\_24 & 0.00839 \\ 
        18\_15 & 0.00544 & 3\_5 & 0.00588 \\ 
        15\_18 & 0.00544 & 5\_3 & 0.00588 \\ 
    \end{tabular*}
\end{table}

\begin{table*}
    \centering
    \caption{Table of the single-qubit device properties of \textit{ibmq\_ehningen} (as of August 18th, 2023). The table displays the frequencies, T1 and T2, gate errors, and readout errors for the various qubits, as reported by IBM.}
    \label{table:backend_properties}
    \begin{tabular*}{0.9 \textwidth}{@{\extracolsep{\fill}}cccccccc}
        Qubit & Frequency / GHz & T1 /µs & T2 /µs & RZ error & SX error & X error& Readout error \\ \hline
        Q0 & 4.961 & 142.59275 & 95.47855 & 0 & 0.00018 & 0.00018 & 0.011 \\
        Q1 & 5.18191 & 181.00386 & 70.72969 & 0 & 0.00021 & 0.00021 & 0.0096 \\
        Q2 & 5.12694 & 95.06903 & 9.06901 & 0 & 0.0003 & 0.0003 & 0.009 \\
        Q3 & 5.26815 & 99.89562 & 28.579 & 0 & 0.00017 & 0.00017 & 0.0229 \\
        Q4 & 5.05357 & 159.58091 & 70.14323 & 0 & 0.00068 & 0.00068 & 0.0123 \\
        Q5 & 5.07116 & 104.48761 & 8.15402 & 0 & 0.00032 & 0.00032 & 0.024 \\ 
        Q6 & 4.89006 & 143.66941 & 122.05597 & 0 & 0.00035 & 0.00035 & 0.014 \\
        Q7 & 4.97776 & 123.99653 & 113.28929 & 0 & 0.00021 & 0.00021 & 0.0076 \\
        Q8 & 5.17419 & 84.87928 & 69.68872 & 0 & 0.00104 & 0.00104 & 0.0245 \\ 
        Q9 & 4.9925 & 112.5169 & 62.87694 & 0 & 0.00573 & 0.00573 & 0.0175 \\ 
        Q10 & 4.83511 & 191.01486 & 162.14782 & 0 & 0.00039 & 0.00039 & 0.0092 \\
        Q11 & 5.11944 & 92.86756 & 127.55782 & 0 & 0.00027 & 0.00027 & 0.017 \\
        Q12 & 4.72549 & 169.03301 & 238.81786 & 0 & 0.00015 & 0.00015 & 0.0117 \\
        Q13 & 4.92598 & 155.88103 & 225.73924 & 0 & 0.00022 & 0.00022 & 0.008 \\ 
        Q14 & 5.17671 & 94.24501 & 258.67336 & 0 & 0.00033 & 0.00033 & 0.0077 \\ 
        Q15 & 4.89299 & 93.54811 & 176.56123 & 0 & 0.00019 & 0.00019 & 0.0105 \\ 
        Q16 & 5.02214 & 194.67907 & 191.30807 & 0 & 0.00028 & 0.00028 & 0.0078 \\ 
        Q17 & 5.13566 & 146.76432 & 22.47776 & 0 & 0.00052 & 0.00052 & 0.0073 \\ 
        Q18 & 4.99642 & 130.38322 & 228.58117 & 0 & 0.00024 & 0.00024 & 0.0128 \\
        Q19 & 4.7841 & 172.28179 & 77.40924 & 0 & 0.00032 & 0.00032 & 0.0137 \\ 
        Q20 & 5.04235 & 204.49692 & 220.82418 & 0 & 0.00053 & 0.00053 & 0.025 \\ 
        Q21 & 4.93974 & 126.28787 & 202.81237 & 0 & 0.00026 & 0.00026 & 0.0076 \\
        Q22 & 4.72513 & 58.42632 & 34.32104 & 0 & 0.00025 & 0.00025 & 0.0131 \\
        Q23 & 4.80479 & 190.68715 & 238.40096 & 0 & 0.00043 & 0.00043 & 0.0084 \\
        Q24 & 5.07449 & 233.74912 & 337.44479 & 0 & 0.00015 & 0.00015 & 0.0075 \\
        Q25 & 4.95019 & 198.39525 & 439.17036 & 0 & 0.00018 & 0.00018 & 0.0076 \\ 
        Q26 & 5.15132 & 190.73366 & 24.68715 & 0 & 0.00015 & 0.00015 & 0.0079 \\
    \end{tabular*}
\end{table*}

\section{\label{sec:transpiled_circuits}Transpiled circuits}
\begin{figure*}
    \centering
    \includegraphics[width=\textwidth]{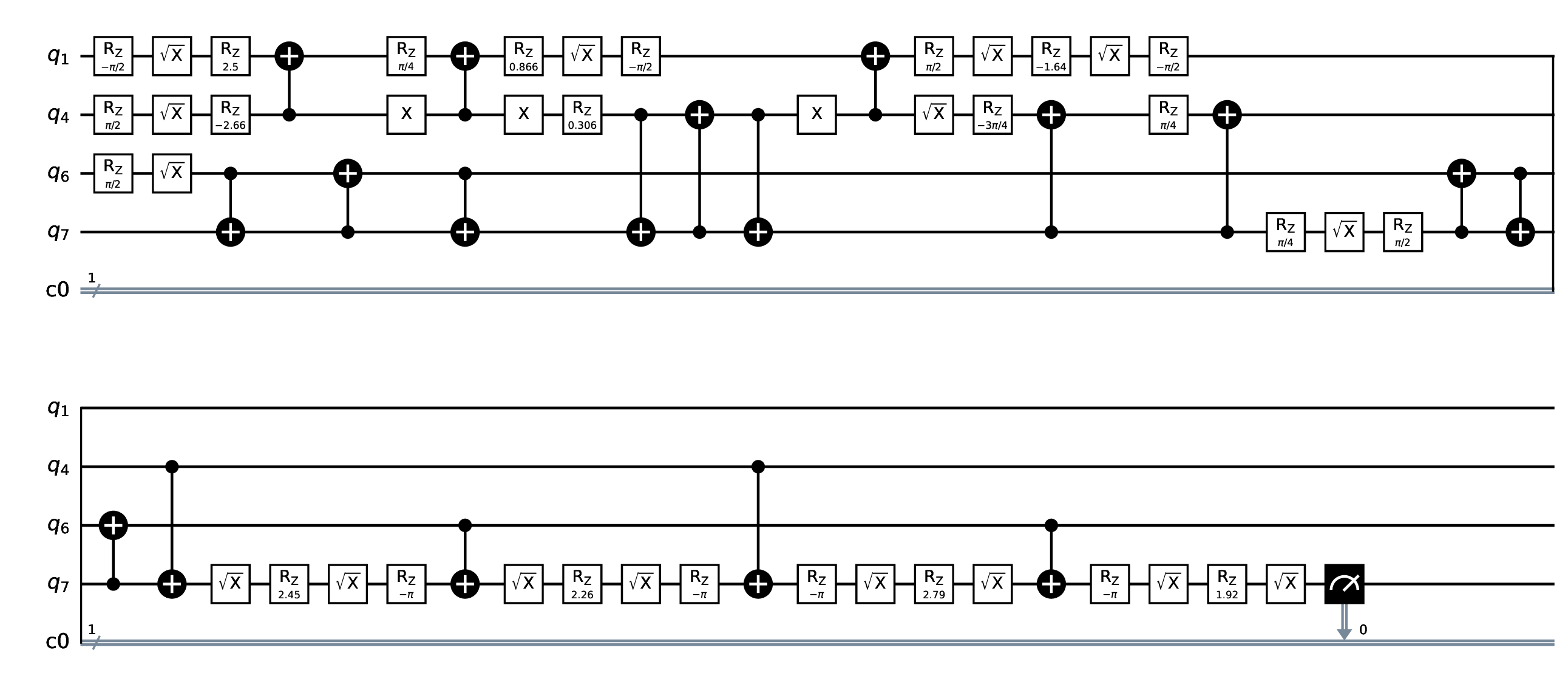}
    \caption{Representation of the HHL circuit which is used in Sec.~\ref{sec:results}. Note that only the last qubit ($q_7$) is measured in case of HHL. The circuit has been designed to conform to IBM's quantum device architecture through transpilation. This requires the use of three SWAP gates, each of which consists of three CNOT gates.}
    \label{fig:qc_t_hhl}
    
    \includegraphics[width=\textwidth]{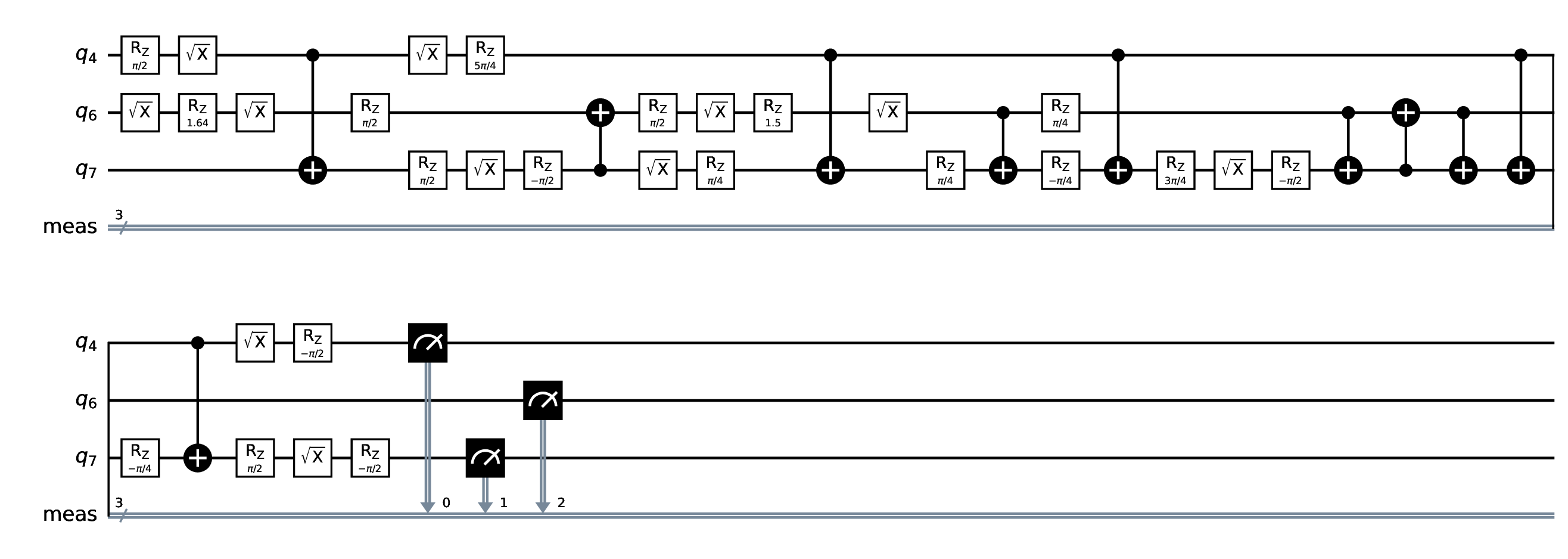}
    \caption{Representation of the Grover circuit which is used in Sec.~\ref{sec:results}. The circuit has been designed to conform to IBM's quantum device architecture through transpilation. This requires the use of one SWAP gate, which consists of three CNOT gates.}
    \label{fig:qc_t_grover}
    
    \includegraphics[width=\textwidth]{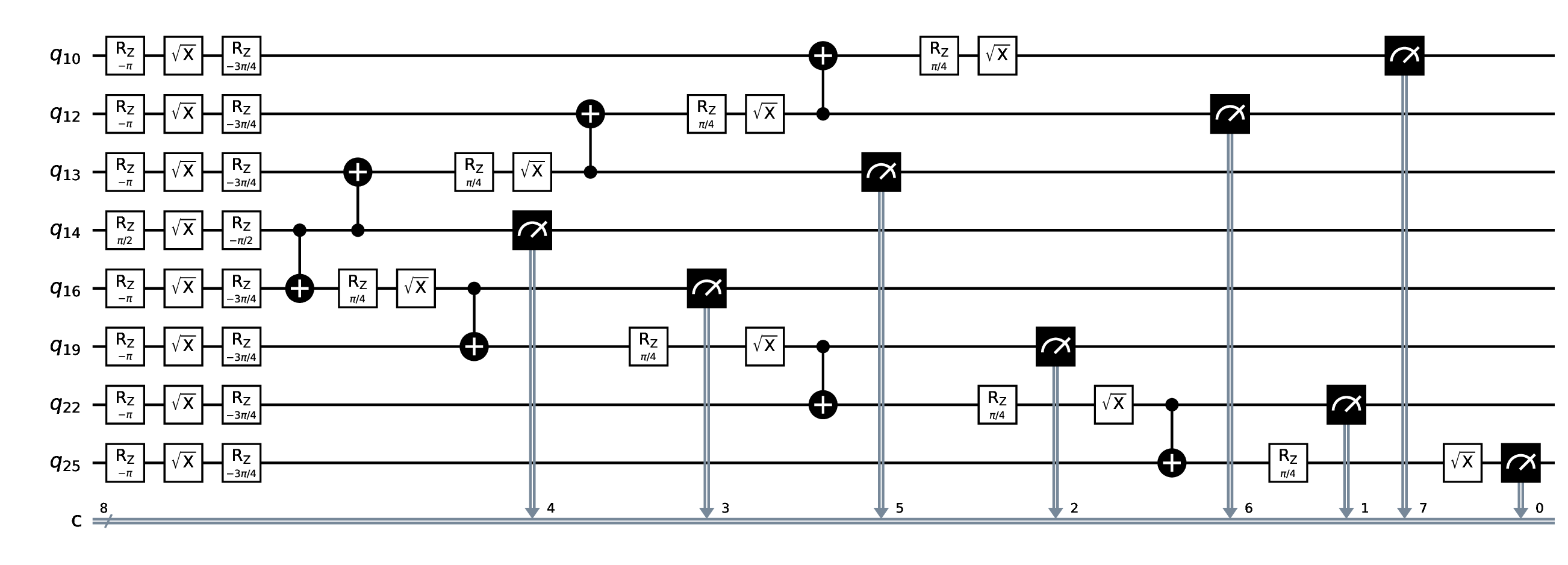}
    \caption{Representation of the Ladder circuit which is used in Sec.~\ref{sec:results}. The circuit has been designed to conform to IBM's quantum device architecture through transpilation.}
    \label{fig:qc_t_ladder}
\end{figure*}

In Figs.~\ref{fig:qc_t_grover}, \ref{fig:qc_t_hhl} and \ref{fig:qc_t_ladder}, we illustrate the quantum circuits utilized in this work, specifically tailored to conform IBM's quantum device architecture through transpilation. The Grover and HHL circuit diagrams incorporate at least on SWAP gate, necessitated by constraints of the device architecture, thereby increasing the number of CNOT gates. The Grover circuit shown in Fig.~\ref{fig:qc_t_grover} requires only one SWAP gate, resulting in a cumulative total of 10 CNOT gates. In contrast, the HHL circuit depicted in Fig.~\ref{fig:qc_t_hhl} exhibits a total of 18 CNOT gates, of which nine are attributed solely to the inclusion of SWAP gates. The H-Ladder circuit does not have additional SWAP gates and the total number of CNOT gates is 7.

\section{\label{sec:asf+iczne} Adaptive scaling factors and filtering with inverted-circuit ZNE}
\begin{figure*}
    \includegraphics[width=\textwidth]{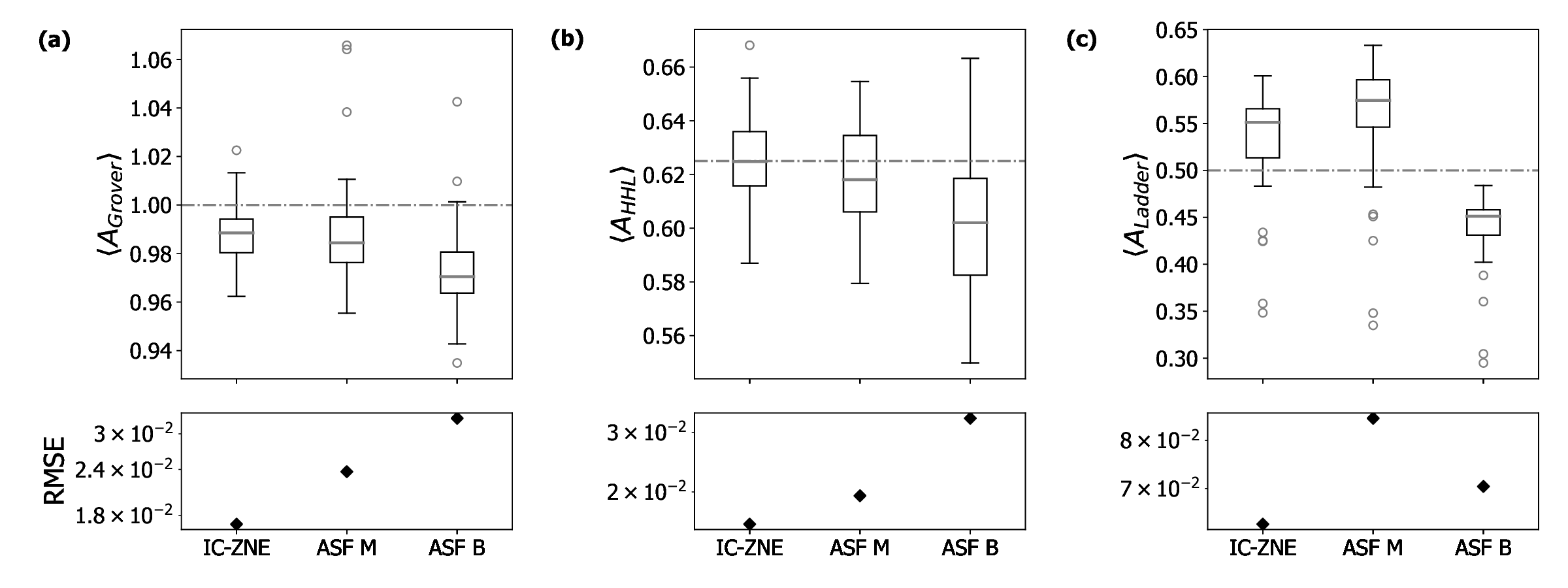}
    \caption{Comparison of the performance of inverted-circuit ZNE (IC-ZNE) for the (a) Grover, (b) HHL and (c) H-Ladder algorithms with adaptive scaling factors (ASF). The exact result is indicated by the dotted horizontal line. The lower subplots depict the root mean square errors (RMSE). Each subplot presents results for IC-ZNE and ASF where $\epsilon_0$ is either estimated from calibration data (ASF M) or measured with an inverted circuit (ASF B).}
    \label{fig:asf+iczne}
\end{figure*}

Adaptive scaling factors (ASF) demonstrated a relative error reduction of up to 23.65\% when combined with standard zero-noise extrapolation (sZNE). However, as discussed in the main text, this improvement does not extend to inverted-circuit zero-noise extrapolation (IC-ZNE). Figure~\ref{fig:asf+iczne} presents the extrapolation results for the Grover (a), HHL (b), and H-Ladder (c) algorithms using IC-ZNE, with the corresponding root mean square error (RMSE) values shown in the lower panels.

Across all three algorithms, the application of ASF in conjunction with IC-ZNE leads to increased RMSE values, indicating a degradation in performance. Notably, for both the Grover and HHL algorithms, ASF values estimated from calibration data outperform those obtained via direct measurement of the initial error strength. This suggests that IC-ZNE already effectively incorporates error information for each data point, thereby reducing or eliminating the benefit of further adjusting the scaling factors. Consequently, ASF provides no further benefit in the context of IC-ZNE and even degrades the quality of the extrapolated results.

Filtering techniques, as outlined in Sec.~\ref{sec:filtering}, can also be extended to the Inverted-Circuit Zero-Noise Extrapolation (IC-ZNE) framework. In this context, we leverage a two-dimensional correlated Gaussian distribution to simultaneously analyze the distribution of expectation values $\langle A \rangle$ and corresponding initial error strengths $\epsilon_0$.

The 2D Gaussian distribution used in this analysis is given by
\begin{widetext}
\begin{align}
f(x, y) = \frac{1}{2\pi \sigma_x \sigma_y \sqrt{1 - \rho^2}} \exp\left(-\frac{1}{2(1 - \rho^2)} \left( \frac{(x - \mu_x)^2}{\sigma_x^2} - \frac{2\rho(x - \mu_x)(y - \mu_y)}{\sigma_x \sigma_y} + \frac{(y - \mu_y)^2}{\sigma_y^2} \right)\right),
\end{align}
\end{widetext}
where $x = \langle A \rangle$ and $y = \epsilon_0$ represent the measured expectation value and initial error strength, respectively. The parameters $\mu_x$, $\mu_y$, $\sigma_x$, and $\sigma_y$ are the mean and standard deviations of $x$ and $y$, and $\rho$ is the Pearson correlation coefficient, defined as $\rho := \frac{\sigma_{x,y}}{\sigma_x \sigma_y}$ with $\sigma_{x,y}$ the covariance between $x$ and $y$.
\begin{figure}
    \centering
    \includegraphics[width=\linewidth]{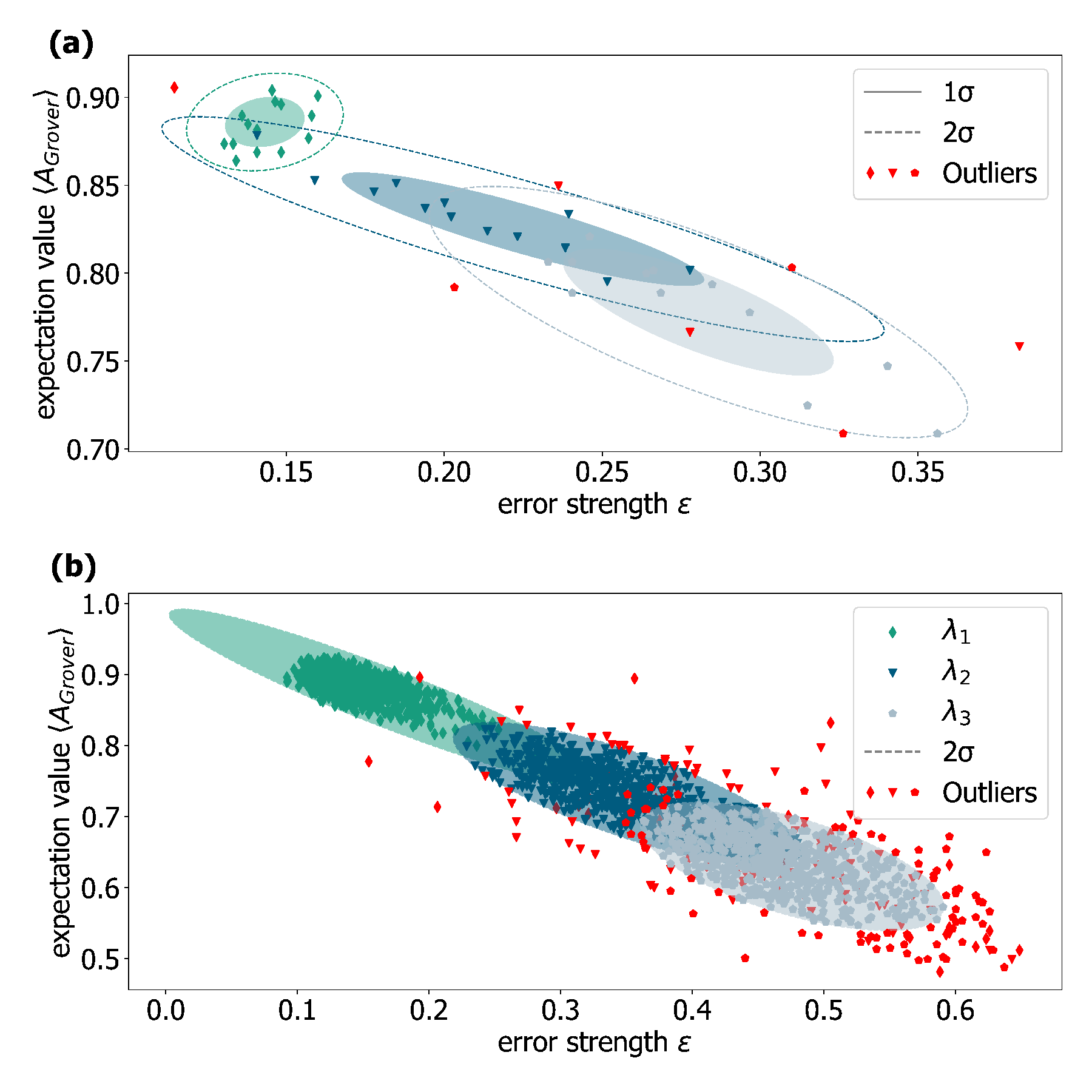}
    \caption{Local and global filtering techniques using IC-ZNE and Grover algorithm. (a) Illustrates local filtering with a 2D Gaussian for each parameter $\lambda_i$, represented by diamonds for $\lambda_1$, triangles for $\lambda_2$, and hexagons for $\lambda_3$. The $1\sigma$ contours are indicated by solid lines, while the $2\sigma$ contours are shown as dotted lines; outliers are highlighted in red. (b) Depicts global filtering across 50 runs utilizing the Grover circuit, with $2\sigma$ regions marked by filled ellipses for each $\lambda_i$. The symbols correspond to those used in (a), and outliers are similarly indicated in red.}
    \label{fig:ic-zne filter}
\end{figure}
The use of a 2D Gaussian distribution enables run-wise filtering by allowing the identification and removal of individual data points that deviate significantly from the joint distribution of expectation value and error strength, a method we refer to as local filtering. In local filtering, illustrated in Fig.~\ref{fig:ic-zne filter}(a), each run is evaluated independently across each noise scaling factor $\lambda_i$. Individual data points are excluded if they lie outside the $2\sigma$ confidence region of the local 2D Gaussian distribution. This approach enables selective removal of anomalous measurements within a single run, providing fine-grained control over the dataset.
\begin{figure}
    \centering
    {\includegraphics[width=0.48\textwidth]{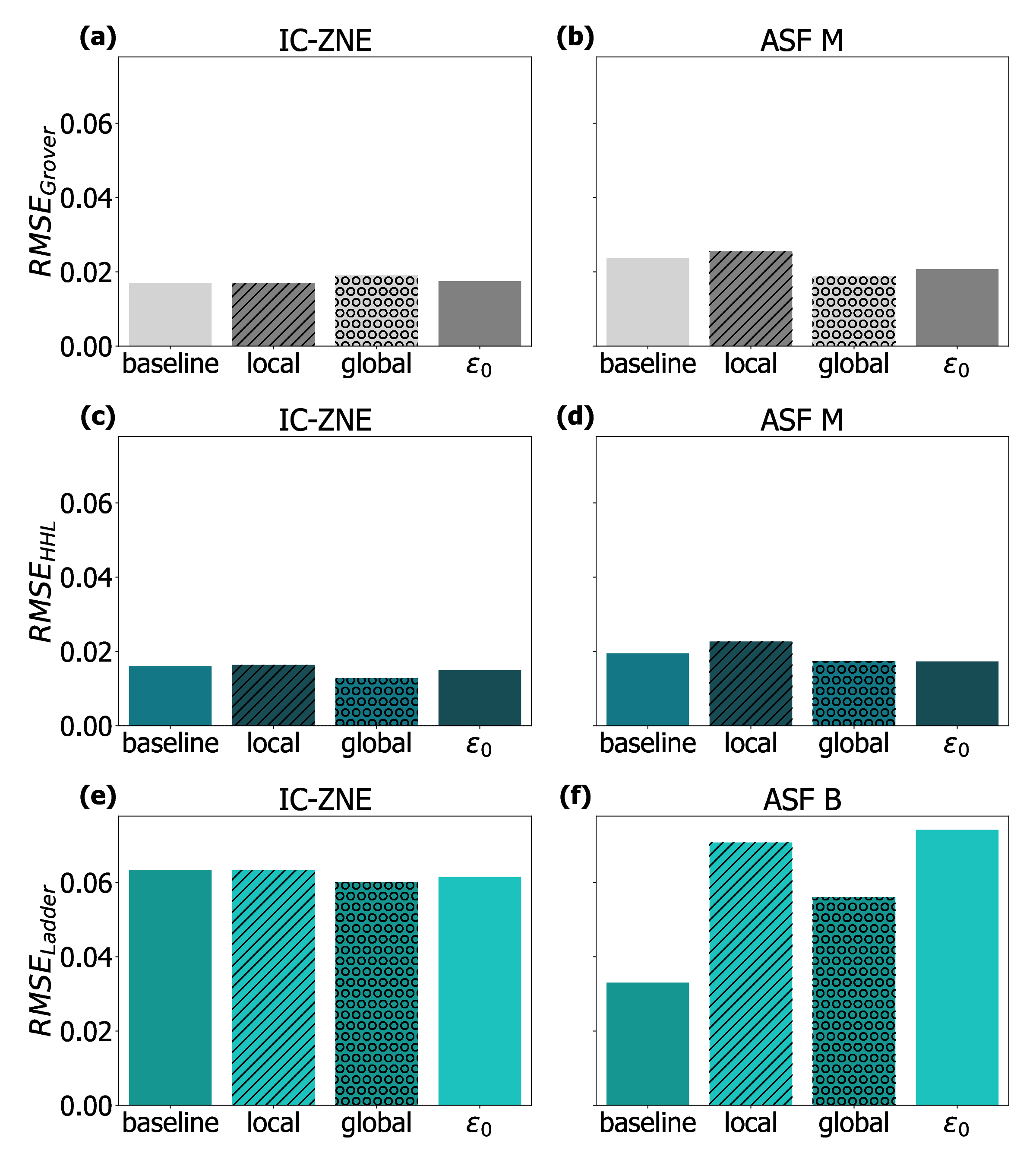}}
    \caption{RMSE by filtering the data by either local, global 
     or $\epsilon_0$ filtering, for the Grover algorithm (a,b), the HHL algorithm (c,d) and the H-Ladder (e,f), using IC-ZNE (a,c,e) or adaptive scaling factors with IC-ZNE (b,d,f). According to Fig.~\ref{fig:asf+iczne}, the better performing variant ASF-M (error strength $\epsilon_0$ estimated from calibration data) is chosen for Grover (b) and HHL (d), and ASF-B ($\epsilon_0$ measured directly on backend) for H-Ladder (f). To enable a direct comparison, the values without filtering are shown (left columns). Second third and fourth column represent local, global and $\epsilon_0$ filtering.}
    \label{fig:RMSE filter iczne}
\end{figure}

In contrast, global filtering (Fig.~\ref{fig:ic-zne filter}(b)) is applied across a larger ensemble of runs, here 50 runs. For each noise scaling factor $\lambda_i$, a global 2D Gaussian distribution is fitted to all corresponding data points collected over multiple experimental realizations. Points outside the $2\sigma$ region of this distribution are discarded, effectively removing systematic deviations that appear across multiple circuits.

While these filtering techniques offer a principled approach to identifying statistical outliers, their application to the circuits studied, does not result in a significant improvement in extrapolation accuracy. Nevertheless, we include these results to highlight the potential utility of the method, especially for circuits or algorithms exhibiting more pronounced or frequent outliers.

Figure~\ref{fig:RMSE filter iczne} compares the RMSE achieved using different filtering strategies—local, global, and $\epsilon_0$-based filtering—for the Grover (a,b), HHL (c,d), and H-Ladder (e,f) algorithms. Each algorithm is evaluated under two extrapolation frameworks: IC-ZNE alone (a,c,e) and IC-ZNE combined with Adaptive Scaling Factors (ASF) (b,d,f). Based on Fig.~\ref{fig:asf+iczne}, the ASF variant is chosen based on best performance: ASF-M (model-based $\epsilon_0$ from calibration data) is used for Grover (b) and HHL (d), while ASF-B (backend-measured $\epsilon_0$) is used for H-Ladder (f). For each setting, the first bar represents the baseline RMSE without filtering, followed by bars corresponding to local, global, and $\epsilon_0$ filtering.

For the Grover algorithm, filtering has little to no effect in the IC-ZNE setting (a), but global filtering provides notable improvement when combined with ASF-M (b).
For HHL, global filtering consistently improves results for both IC-ZNE and ASF-M (c,d), likely due to the presence of distinct outliers that are effectively removed.
For the H-Ladder algorithm, IC-ZNE benefits slightly from both global and $\epsilon_0$ filtering (e), but when combined with ASF-B (f), filtering introduces greater variability and leads to worse overall performance.

\end{document}